\documentclass[aps,prl,superscriptaddress,showpacs,twocolumn,footinbib]{revtex4-1}
\usepackage[pdftex]{graphicx}
\usepackage{amsmath}
\usepackage{xspace}
\usepackage{color}
\usepackage[colorlinks=true,linkcolor=blue,urlcolor=blue,citecolor=blue]{hyperref}
\usepackage[normalem]{ulem}
\usepackage{hyperref}

\usepackage{color}
\definecolor{BV}{rgb}{0.1,0.,0.6}
\definecolor{R}{rgb}{0.9,0,0}
\definecolor{G}{rgb}{0.2,0.8,0.2}
\usepackage{comment}

\begin{document}

\title{The impact of the injection protocol on an impurity's stationary state}

\author{Oleksandr Gamayun}
\affiliation{Instituut-Lorentz, Universiteit Leiden, P.O. Box 9506, 2300 RA Leiden, The Netherlands}
\affiliation{Bogolyubov Institute for Theoretical Physics, 14-b Metrolohichna str., Kyiv 03680, Ukraine}

\author{Oleg Lychkovskiy}

\affiliation{Skolkovo Institute of Science and Technology,
Skolkovo Innovation Center 3, Moscow  143026, Russia}

\affiliation{Steklov Mathematical Institute of Russian Academy of Sciences,
Gubkina str. 8, Moscow 119991, Russia}

\affiliation{Russian Quantum Center, Novaya St. 100A, Skolkovo, Moscow Region, 143025, Russia}

\author{Evgeni Burovski}

\affiliation{National Research University Higher School of Economics, 101000 Moscow, Russia}
\affiliation{Science Center in Chernogolovka,142432 Chernogolovka, Russia}

\author{Matthew Malcomson}
\affiliation{Physics Department, Lancaster University, Lancaster LA1 4YB, United Kingdom}

\author{Vadim V. Cheianov}
\affiliation{Instituut-Lorentz, Universiteit Leiden, P.O. Box 9506, 2300 RA Leiden, The Netherlands}

\author{Mikhail B. Zvonarev}
\affiliation{LPTMS, CNRS, Univ. Paris-Sud, Universit\'e Paris-Saclay, 91405 Orsay, France}

\date{\today}

\begin{abstract}

We examine stationary state properties of an impurity particle injected into a one-dimensional quantum gas. We show that the value of the impurity's end velocity lies between zero and the speed of sound in the gas, and is determined by the injection protocol. This way, the impurity's constant motion is a dynamically emergent phenomenon whose description goes beyond accounting for the kinematic constraints of Landau approach to superfluidity.  We provide exact analytic results in the thermodynamic limit, and perform finite-size numerical simulations to demonstrate that the predicted phenomena are within the reach of the existing ultracold gases experiments.

\end{abstract}

\maketitle

Understanding and controlling the propagation of a particle in a medium is a basic problem of physics, with applications ranging from neutron moderation to the design of semiconductor heterostructures~\cite{kuper_book, prokofev_diffusion_impurity_93, devreese_polaron_09}. Experimental realizations of two-component mixtures of ultracold atomic gases with large concentration imbalance offer new perspectives for this problem. In particular, they give access to systems with reduced spatial dimensions, in the continuum~\cite{palzer_impurity_transport_09, catani_impurity_dynamics_11, meinert_bloch_16} as well as on a lattice~\cite{fukuhara_spin_impurity_2013}. Cooled down to virtually zero temperature, such systems demonstrate a remarkably non-trivial interplay of the effects due to quantum statistics and strong correlations. In particular, it was predicted recently that the velocity of an impurity injected into the gas may experience underdamped oscillations around some stationary-state value, a phenomenon called the quantum flutter~\cite{mathy_flutter_2012, knap_flutter_signatures_2014}. A possibility of the impurity's constant motion could be explained within Landau approach to superfluidity, based on the kinematic restrictions for the possible outcomes of the impurity-gas scattering~\cite{lychkovskiy_perpetual_2014, lychkovskiy_perpetual_15}. It can also be seen by solving a quantum Boltzmann equation obtained from perturbation theory for a small impurity-gas interaction strength~\cite{burovski_impurity_momentum_2014, gamayun_kinetic_impurity_TG_14,gamayun_quantum_boltzmann_14}.

In order to characterize the impurity stationary state properties under realistic experimental conditions, two challenging problems have to be resolved on the theory side. First, one has to learn how to deal with a finite impurity-gas interaction which is strong enough to render any existing perturbation and quantum Boltzmann theory inapplicable~\footnote{Strong impurity-gas interaction is required to ensure a sufficiently large number of the scattering events while the impurity goes through a finite-size atomic cloud, like in the experiment~\cite{meinert_bloch_16}}. Second, one has to identify the dependence of the stationary state properties on how the system is prepared; in particular, on the rate the impurity-gas interaction is turned on. One commonly used approach to these theoretical problems is  to simulate real-time dynamics numerically.  This has been done by either using a selected subset of the exact eigenstates known from the Bethe ansatz~\cite{mathy_flutter_2012,robinson_impurity_16} or by using the time-dependent density-matrix renormalization group~\cite{peotta_mobile_impurity_TDMRG_2013,massel_mobile_impurity_TDMRG_2013,knap_flutter_signatures_2014}. Such simulations are limited to a finite time domain, and a few tens of particles. However, the effects from the finite particle number can be very significant~\cite{altland_Landau-Zener_08,altland_nonadiabaticity_09}, and the time to reach the stationary state may not be numerically accessible. An analytic solution could be largely simplified by a polaron approximation, whose drawback is a limited applicability range~\cite{catani_impurity_dynamics_11, peotta_mobile_impurity_TDMRG_2013, bonart_impurity_luttinger_13, massel_mobile_impurity_TDMRG_2013}.

\begin{figure*}[ht]
\centering
\includegraphics[width=0.98\linewidth, clip=true,trim= 0 0 0 0]{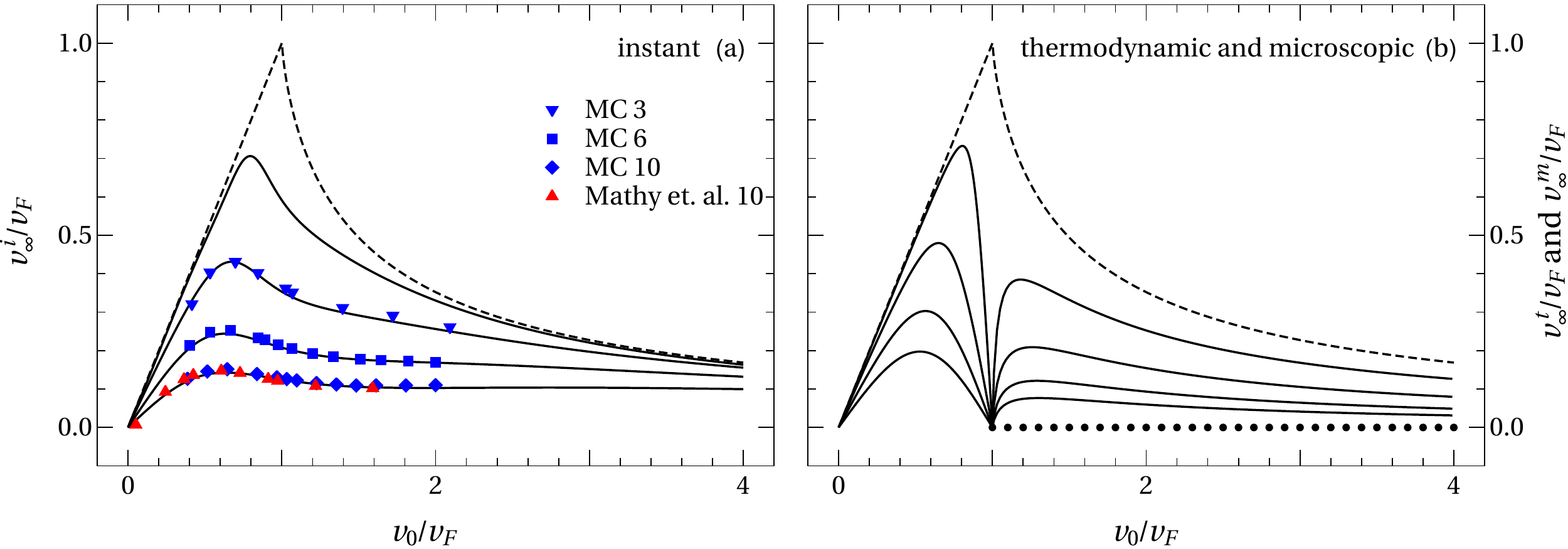}
\caption{(Color online)
Impurity's end velocity shown as a function of the impurity's initial velocity $v_0$. Left panel is for the instant injection protocol. Solid curves illustrate our analytic results for the impurity-gas interaction strength $\gamma=1,$ $3,$ $6,$ and $10$ (top to bottom). Blue down-triangles, boxes, and diamonds come from the stochastic Monte-Carlo simulations for $\gamma=3,$ $6,$ and $10$, respectively. Red up-triangles are the results for $\gamma=10$ obtained by combining Bethe ansatz and numerics in Ref.~\cite{mathy_flutter_2012}.  Dashed curve corresponds to $v_\infty^i$ in the $\gamma\to 0$ limit. All curves converge to the $\gamma$-independent asymptotic form $v_\infty^i=2v_F^2/(3v_0)$ in the large $v_0$ limit. Right panel is for the two adiabatic injection protocols: thermodynamic ($v_\infty^{t}$, solid lines) and microscopic ($v_\infty^{m}$, dotted lines). The values of $\gamma$ are the same as on the left panel. These two protocols lead to the same value of the end velocity of the impurity, $v_\infty^{t} = v_\infty^m$, for $v_0<v_F$, and therefore dotted lines are not plotted on top of the solid ones. However, $v_\infty^t$ is different from $v_\infty^m$ for $v_0>v_F$, where $v_\infty^{m}=0$ for all values of $\gamma$. The $\gamma\to 0$ limit of $v_\infty^t$ tends to the same dashed curve as in the left panel for all $v_0\ne v_F$, and $v^t_\infty=0$ for $v_0=v_F$. The units are set by the Fermi velocity $v_F=\pi\rho/m$. }
\label{fig:pinfvsq}
\end{figure*}

In this Letter, we investigate stationary state properties of an impurity particle injected with some initial velocity $v_0$ into a one-dimensional Fermi gas. We characterize the impurity injection protocol by the rate at which the impurity-gas interaction constant $\gamma$ is turned on. We identify three protocols. First, instant injection, corresponds to ramp-ups of $\gamma$ much larger than the Fermi energy $E_F$. Second, microscopically adiabatic injection, is an opposite limit of the instant one. It corresponds to ramp-ups much smaller than the level spacing at a given energy. This condition gets more restrictive as the number of particles in the gas increases. Third, thermodynamically adiabatic injection, is an intermediate case, corresponding to ramp-ups much larger than the level spacing but much smaller than $E_F$. These protocols lead to qualitatively different behavior of the impurity's end velocity. In particular, in the case of the instant protocol the end velocity is non-zero for any $v_0$; in the case of the microscopically adiabatic one the end velocity vanishes for any $v_0$ larger than the Fermi velocity $v_F$; and finally, in the case of the thermodynamically adiabatic one the end velocity vanishes only at $v_0=v_F$. We illustrate the end velocity of the impurity in Fig.~\ref{fig:pinfvsq}, derived from our exact analytic solution in the thermodynamic limit. We perform large-scale stochastic Monte Carlo simulations to show that there is no visible deviation from our predictions due to finite-size effects for experimentally relevant parameters of ultracold atomic gases.

The Hamiltonian of our system, consisting of an impurity of mass $m$ interacting with the gas particles of the same mass via a repulsive $\delta$-function potential reads
\begin{equation}
H= \sum_{j=1}^N \frac{ P_j^2}{2m} + \frac{ P_\mathrm{imp}^2}{2m}+ \frac{\gamma\rho}{m} \sum_{j=1}^N \delta(x_j-x_\mathrm{imp}). \label{eq:Htot}
\end{equation}
Here, $x_j$ ($ P_j$) is the coordinate (momentum) of the $j$th gas particle, $j=1,\ldots,N,$ and $x_\mathrm{imp}$ ($P_\mathrm{imp}$) is that of the impurity. Planck's constant $\hbar=1$ in our units, $\gamma$ represents the dimensionless strength of the impurity-gas repulsion, $\rho=N/L$ is the gas density, and $L$ is the system size. Our analytic results are obtained in the thermodynamic limit of large $N$ and $L$ at a fixed $\rho$ and zero temperature. We start the system out in the state
\begin{equation}
|\mathrm{FS},Q \rangle = c^\dagger_{\mathrm{imp}}(Q)|0\rangle \otimes |\mathrm{FS}\rangle, \label{eq:inQ}
\end{equation}
where $c^\dagger_{\mathrm{imp}}(Q)$ creates an impurity plane wave with momentum  $Q=mv_0$ from the vacuum $|0\rangle$, and a free spinless Fermi gas is in the Fermi sea ground state $|\mathrm{FS}\rangle $. We are interested in the impurity's velocity at infinite time
\begin{equation}
v_\infty = \frac1{m} \lim_{t\to\infty} \langle \mathrm{FS},Q|P_\mathrm{imp}(t)|\mathrm{FS},Q \rangle, \label{eq:viadef}
\end{equation}
where $P_\mathrm{imp}(t)$ is the impurity momentum operator in the Heisenberg representation.

We begin by considering an instant quench of the impurity-gas interaction. The time evolution of $P_\mathrm{imp}$ is determined by the Hamiltonian~\eqref{eq:Htot} with a given interaction strength $\gamma$:
\begin{multline}
v^i(t) =\frac{1}{m} \sum_{f_\gamma, f^\prime_\gamma} e^{-it(E_{f_\gamma} - E_{f^\prime_\gamma})}\\
\times\langle f^\prime_\gamma|P_\mathrm{imp} |f_\gamma\rangle \langle  \mathrm{FS},Q |f^\prime_\gamma  \rangle \langle f_\gamma | \mathrm{FS},Q \rangle.
\label{eq:vt}
\end{multline}
The double sum runs over complete sets of the eigenfunctions $|f_\gamma\rangle$ and $|f^\prime_\gamma\rangle$ of the Hamiltonian~\eqref{eq:Htot} having total momentum $Q$; $E_{f_\gamma}$ and $E_{f^\prime_\gamma}$ stand for the energies of these states. We denote the velocity from Eq.~\eqref{eq:viadef} obtained for the instant injection protocol as $v_\infty^i$. We have 
\begin{equation}
v_\infty^i= \lim_{t\to\infty} v^i(t) =\frac{1}{m} \sum_{f_\gamma} \langle f_\gamma|P_\mathrm{imp} |f_\gamma\rangle |\langle f_\gamma | \mathrm{FS},Q \rangle|^2.
\label{eq:visum}
\end{equation}
Here, we reduced the double sum from Eq.~\eqref{eq:vt} to the single sum by assuming that only the terms with $E_{f_\gamma} = E_{f^\prime_\gamma}$ are relevant in the $t\to\infty$ limit. The states $|f_\gamma\rangle$ are found exactly by the Bethe ansatz technique for any finite $N$ and periodic boundary conditions~\cite{mcguire_impurity_fermions_65}. Note that Eq.~\eqref{eq:Htot} is a particular case of the Gaudin-Yang model~\cite{gaudin_fermions_spinful_67, yang_fermions_spinful_67}. Evaluating the sum in Eq.~\eqref{eq:visum} poses a separate challenge. We do this following the summation procedure developed in Refs.~\cite{gamayun_impurity_Green_FTG_14, gamayun_impurity_Green_FTG_16}, aiming at the thermodynamic limit, where boundary conditions play no role. The matrix elements $\langle f_\gamma|P_\mathrm{imp} |f_\gamma\rangle$ turn into an analytic function $\mathcal{P}(\Lambda)$ of a single argument $\Lambda$ in the thermodynamic limit. This way the sum in Eq.~\eqref{eq:visum} is expressed as the double integral~\cite{Supp},
\begin{equation}
\frac{v_\infty^i}{v_F} =  -i \int_{-\infty}^\infty \frac{d\Lambda}{\pi} \mathcal{P}(\Lambda) \int_0^\infty dx\, \sin(xv_0/v_F) F(\Lambda,x). \label{eq:Pinf_exact}
\end{equation}
Here,
\begin{equation}
\mathcal{P}(\Lambda)  =  \frac{\Lambda }{\alpha }+\frac{1}{2\alpha}\frac{\ln\frac{1+(\alpha -\Lambda )^2}{1+(\alpha +\Lambda )^2}}{\arctan(\alpha -\Lambda) + \arctan(\alpha +\Lambda)}, 
\end{equation}
where $\alpha = 2\pi/\gamma$, and $k_F=\pi\rho$ and $v_F=k_F/m$ stands for the Fermi momentum and velocity, respectively. The function $F$ reads
\begin{equation}
F(\Lambda,x)=(h-1)\det(\hat I + \hat V) + \det(\hat I + \hat V - \hat W). \label{eq:F}
\end{equation}
The ``det'' symbol  stands for the Fredholm determinant of the linear integral operators $\hat V$ and $\hat W$ defined on the domain $[-1,1]\times[-1,1]$ with the following kernels:
\begin{equation}
V(q,q^\prime) = \frac{e_+(q)e_-(q^\prime)-e_-(q)e_+(q^\prime)}{q-q^\prime}
\end{equation}
and
\begin{equation}
W(q,q^\prime) = e_+(q) e_+(q^\prime).
\end{equation}
The functions $e_\pm$ are defined as
\begin{equation}
e_+(q) =\frac{e(q)e^{-ixq/2}}{\sqrt{\pi}}, \quad e_-(q) = \frac{e^{-ixq/2}}{\sqrt{\pi}},
\end{equation}
where
\begin{equation}\label{eq:exact}
e(q) = - \frac{ e^{iq x}- \alpha h}{\Lambda+i-\alpha q}, \quad h =\frac{1}{\alpha} e^{ix (\Lambda+i)/\alpha}.
\end{equation}

Let us discuss the impurity's stationary state in the $\gamma\to0$ limit. Equation~\eqref{eq:Pinf_exact} implies~\cite{Supp}
\begin{equation}
v_\infty^i = v_0 - \theta(v_0-v_F) \frac{v_0^2-v_F^2}{2v_F} \ln\frac{v_0+v_F}{v_0-v_F}, \quad \gamma\to 0. \label{eq:vigto0}
\end{equation}
Here, $\theta$ is the Heaviside step function. This expression, shown with the dashed line in Fig.~\ref{fig:pinfvsq}, has already been obtained in Ref.~\cite{burovski_impurity_momentum_2014}, by evaluating the sum in Eq.~\eqref{eq:visum} with a technique special to the $\gamma\to 0$ limit. We stress that this limit is taken assuming the stationary state is already reached, which probably takes very long time. Thus, however small but finite $\gamma$ leads to impurity-gas scattering processes significantly modifying the stationary state of the impurity compared to the initial state~\eqref{eq:inQ} if $v_0>v_F$, and this does not happen for $v_0<v_F$.
The density matrix $\varrho_\infty^i$ characterizing the impurity in the stationary state has not been found yet even in the $\gamma\to 0$ limit. We assume that the properties of this matrix are not affected by excitations in the gas caused by the equilibration process. Exploiting this assumption, we take $\varrho_\infty^i$ as a statistical mixture of the minimal energy states, which are given by $|\mathrm{FS},k\rangle$ with the total momentum lying in the interval $-k_F<k<k_F$:
\begin{equation}
\varrho_\infty^i = \sum_{|k|<k_F} n_k(Q) |\mathrm{FS},k\rangle \langle \mathrm{FS},k|, \quad \gamma\to 0.
\label{eq:rho}
\end{equation}
The coefficients $n_k$ determine the impurity's momentum distribution; they can be taken from Ref.~\cite{gamayun_quantum_boltzmann_14}:
\begin{equation}
n_k(Q) =  
\delta_{k,Q} + \frac{2\pi}L \frac{Q^2-k_F^2}{2k_F} \frac{\theta(Q-k_F)}{(k-Q)^2}  
\label{eq:nq}
\end{equation}
for $-k_F \le k \le k_F$, and vanish outside this interval. Thus, Eq.~\eqref{eq:rho} bridges exact many-body quantum mechanics and statistical physics, the latter not requiring full knowledge of the system's dynamics for a description of a local microscopic object. We later use this equation to characterize the impurity injected adiabatically slow.

We now further examine the dependence of $v_\infty^i$ on the initial velocity. The large $v_0$ limit of Eq.~\eqref{eq:Pinf_exact} reads~\cite{Supp}
\begin{equation}
v_\infty^i = \frac{2}3 \frac{v_F^2}{v_0}, \qquad v_0 \gg v_F.
\label{eq:PinfQlarge}
\end{equation}
One can see from this expression that $v_\infty^i$ decays with increasing $v_0$, which may seem paradoxical. To understand such a behavior 
recall that a single scattering event of an impurity with momentum $Q$ and a gas particle with momentum $k$ yields the impurity's reflection probability $R(k)=\gamma^2 \rho^2/[(k-Q)^2+\gamma^2 \rho^2]$. A toy model accounting for a single collision with gas particles having the Fermi sea momentum distribution, and ignoring that the time until such a collision takes place depends on the gas particle's momentum, gives for the average impurity's velocity
\begin{equation}
\tilde v^i = \frac1{m}\frac{\int_{-k_F}^{k_F} dk\, k R(k)}{\int_{-k_F}^{k_F} dk\, R(k)}. \label{eq:tildePinf}
\end{equation}
This formula correctly reproduces Eqs.~\eqref{eq:vigto0} and~\eqref{eq:PinfQlarge}, illustrating that the dependence of the impurity's end velocity on the initial one is non-monotonous.

For the intermediate values of $v_0$ and $\gamma$ we evaluate the Fredholm determinants entering Eq.~\eqref{eq:Pinf_exact} numerically~\cite{Supp} using very efficient numerical procedure described in Ref.~\cite{bornemann_10}. The resulting plots are shown in Fig.~\ref{fig:pinfvsq}(a) with the solid lines. We then verify that the time-dependent impurity velocity~\eqref{eq:vt} found for a finite particle number $N$ numerically converges to $v_\infty^i$ given by Eq.~\eqref{eq:Pinf_exact} with increasing $t$ and $N$. We tackle the sum in Eq.~\eqref{eq:vt} by the stochastic enumeration method~\cite{burovski_impurity_momentum_2014, malcomson_16}. It constructs a random walk in the space of the Bethe ansatz states $|f_\gamma\rangle$ based on the Metropolis algorithm~\cite{metropolis_53} with the Monte Carlo weight given by $|\langle f_\gamma | \mathrm{FS},Q \rangle|^2$, thus finding the most relevant states automatically. The results shown with the blue down-triangles, boxes, and diamonds in Fig.~\ref{fig:pinfvsq}(a) are for $N=99$. The error bars are smaller than the size of the symbols; the positions of the symbols do not visibly change with further increase of $N$~\footnote{When applied to Eq.~\eqref{eq:visum}, the stochastic enumeration method makes it possible to tackle a system containing up to $400$ particles.}. The deterministic truncation of the sum in Eq.~\eqref{eq:vt} only retaining terms with up to three quasi-particle-hole excitations, employed in Ref.~\cite{mathy_flutter_2012}, leads to the results shown with the red up-triangles.

We now discuss the adiabatic injection protocols mentioned in the introduction. An important feature of the initial state~\eqref{eq:inQ} is its overlap with an eigenstate $|\tilde f_\gamma\rangle$ of the Hamiltonian~\eqref{eq:Htot} at a given total momentum $Q$~\cite{Supp}:
\begin{equation}
\langle \mathrm{FS},Q|\tilde f_\gamma\rangle \to 1, \quad N\to\infty \text{ after } \gamma\to0.
\label{eq:ov}
\end{equation}
Assuming that the ramp-up of $\gamma$ is sufficiently small for the system to stay close to $|\tilde f_\gamma\rangle$ at all times, we get for Eq.~\eqref{eq:viadef}
\begin{equation}
v_\infty^m =\frac1{m} \langle \tilde f_\gamma|P_\mathrm{imp}| \tilde f_\gamma\rangle.
\label{eq:vgsd}
\end{equation}
The superscript ``$m$'' stands for microscopic, indicating that the above assumption exploits the structure of the excitation spectrum at the microscopic level. Using the Hellmann-Feynman theorem as explained in Ref.~\cite{knap_flutter_signatures_2014} we get for Eq.~\eqref{eq:vgsd}
\begin{equation}
v_\infty^m(Q) = \frac{\partial E_{\tilde f_\gamma} (Q)}{\partial Q},
\label{eq:vafpf}
\end{equation}
where $E_{\tilde f_\gamma}(Q)$ is the energy of the state $|\tilde f_\gamma \rangle$.

\begin{figure}
\centering
\includegraphics[width=0.98\linewidth, clip=true,trim= 0 0 0 0]{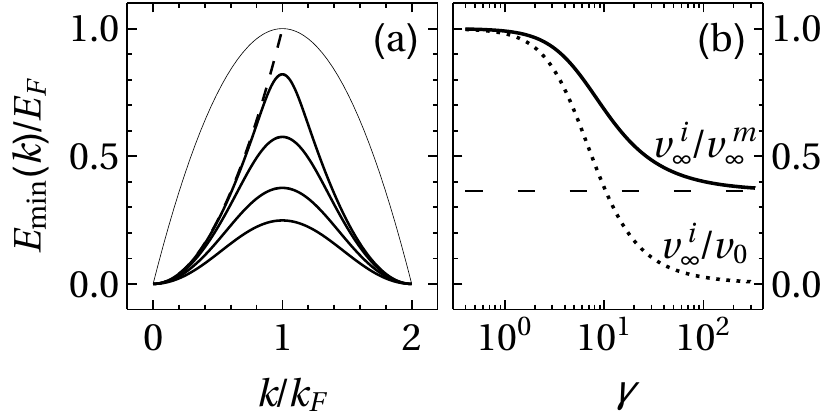}
\caption{(a) The minimum of the excitation spectrum $E_\mathrm{min}(k)$ relative to $E_\mathrm{min}(0)$ for the impurity-gas interaction strength $\gamma=1,$ $3,$ $6,$ and $10$ (thick solid lines, in the order of decreasing magnitude). Dashed line corresponds to $E_\mathrm{min}(k)$ for $0\le k \le k_F$ at $\gamma=0$, which is the impurity's kinetic energy. Thin solid line represents the minimum of the excitation spectrum of the free Fermi gas. (b) $v_\infty^i/v_0$ (dotted line) and $v_\infty^i/v_\infty^m$ (solid line) in the $v_0 \to 0$ limit. Dashed horizontal line indicates $v_\infty^i/v_\infty^m=0.364\ldots$ in the $\gamma\to\infty$ limit.}
\label{fig:d}
\end{figure}

In the case $v_0<v_F$ the state $|\tilde f_\gamma\rangle$ minimizes the Hamiltonian~\eqref{eq:Htot}  at a given $Q$, and is non-degenerate. Hence, Eq.~\eqref{eq:ov} holds for arbitrary $N$. The energy $E_{\tilde f_\gamma}$, denoted as $E_\mathrm{min}$, is shown with thick solid lines in Fig.~\ref{fig:d}(a) for several values of $\gamma$. This energy is quadratic for small $Q$, and $v_\infty^m=v_0 m/m_*$ is the velocity of a particle-like excitation, a polaron with an effective mass $m_*$. We then compare $v_\infty^i$ and $v_\infty^m$ as $v_0\to 0$ and demonstrate in Fig.~\ref{fig:d}(b) that $v_\infty^i<v_\infty^m<v_0$ for any $\gamma>0$~\cite{Supp}.  Thus, the stationary state formed past an instant injection with $v_0\to 0$ can not be described within the polaron theory approach suitable for the state formed past an adiabatic injection. 


In the case $v_0>v_F$ the initial state~\eqref{eq:inQ} is a degenerate non-minimal eigenstate of the Hamiltonian~\eqref{eq:Htot} at $\gamma=0$, and Eq.~\eqref{eq:ov} holds only in the $N\to\infty$ limit. We identify $|\tilde f_\gamma \rangle$ rigorously by examining the Bethe ansatz solution, and Eq.~\eqref{eq:vafpf} gives~\cite{Supp}
\begin{equation}
v_\infty^m= 0, \quad v_0> v_F
\label{eq:vafpf>}
\end{equation}
regardless of the value of $\gamma$. We thus found that sufficiently slow ramp-ups of the interaction lead to a complete stop of the impurity initiated in the state~\eqref{eq:inQ} whose energy is above the minimal excitation energy of a free Fermi gas by a finite amount.
 
The thermodynamically large system has vanishing level spacing. This poses the problem of the validity of Eq.~\eqref{eq:vgsd} since arbitrarily slow ramp-up of the interaction strength could possibly drive the system away from the state $|\tilde f_\gamma \rangle$. Let us impose a thermodynamic adiabaticity condition, more relaxed than the microscopic one. It is realized if the rate of change of the interaction constant $\gamma$ is smaller than the macroscopic energy scales of the system. We then exploit an assumption of separation of time-scales. The change of $\gamma$ from zero to a small finite value is viewed as an instant quench driving the system away from the state $|\tilde f_{\gamma=0} \rangle$, though being slow enough to let the impurity reach the stationary state given by the density matrix~\eqref{eq:rho}. Each term of $\varrho_\infty^i$ is the projector onto a state given by Eq.~\eqref{eq:inQ}, with the momentum $|k|<k_F$. A subsequent slow change of $\gamma$ is assumed to make each of those states follow its own path of the microscopic adiabatic evolution, and we get for the end velocity of the impurity in the thermodynamically adiabatic injection protocol
\begin{equation}
v_\infty^t(Q) = \sum_{|k|<k_F} n_k(Q) v_\infty^m(k),  \label{eq:va2}
\end{equation}
where $n_k$ is given by Eq.~\eqref{eq:nq}, and $v_\infty^m$ by Eq.~\eqref{eq:vafpf}. One can see that $v_\infty^t$ and $v_\infty^i$, illustrated in Fig.~\ref{fig:pinfvsq}, only coincide in the $\gamma\to 0$ and $\gamma\to\infty$ limits. Note also, that for any $v_0$, $v_\infty^i$  is smaller than the maximum of $v_\infty^t$ with respect to $Q$. This proves a cojecture from Ref.~\cite{knap_flutter_signatures_2014}.

We recapitulate that the difference between the microscopically and thermodynamically adiabatic injection protocols can be seen by comparing $v_\infty^m$ and $v_\infty^t$, given by Eqs.~\eqref{eq:vgsd} and~\eqref{eq:va2}, respectively. Such a difference arises for $v_0>v_F$, in which case $v_\infty^m$ vanishes, Eq.~\eqref{eq:vafpf>}, and $v_\infty^t$ is non-zero, as shown in Fig.~\ref{fig:pinfvsq}(b). Determining the rate of change of $\gamma$ for which a transition from $v_\infty^m$ to $v_\infty^t$ happens in a large but finite system is a challenging open problem. The density of states of a free Fermi gas grows exponentially with $N$ for any value of the energy that is above the minimum of the excitation spectrum by a finite amount~\cite{bohr_book}. Given such a rapid growth, this rate probably diminishes very fast with $N$. However, existing quantitative theories of adiabaticity breaking are not based on the properties of the density of states~\cite{lychkovskiy_timescale_16,lychkovskiy_timescale_17}, and their implementation for our problem requires a separate study.

In summary, we performed a quantitative study of the impurity's stationary state for one-dimensional Fermi gas (the results for the Tonks-Girardeau gas of strongly repulsive bosons will be identical~\cite{mathy_flutter_2012}). Our analysis shows that (i) The stationary state of an impurity moving through the gas is not completely determined by the kinematic constraints of Landau approach to superfluidity. In particular, the value of the impurity's end velocity depends on the initial velocity $v_0$, as well as on how the impurity-gas interaction strength $\gamma$ is ramped up with time. We demonstrated that a quantitative study is necessary to understand when this value is zero. In other words, neither zero nor non-zero value of the impurity's end velocity seem to be protected by symmetry or kinematic constraints in case of a general injection protocol. Note that our results are consistent with the assumption that the Landau critical velocity (that is, the velocity of sound, equal to $v_F$ for our model) is an upper bound for the impurity's end velocity. (ii) If the impurity's initial kinetic energy is above the minimum of the excitation spectrum of the Fermi gas, that is $v_0 >v_F$, the microscopic adiabaticity condition (the ramp-up of $\gamma$ is smaller than the level spacing) and the thermodynamic one (the ramp-up of $\gamma$ is larger than the level spacing but smaller than the Fermi energy) lead to different impurity's steady states and end velocities. Since (i) and (ii) are qualitative statements, they hold for one-dimensional gases with arbitrary interactions. Finally, our numerical simulations for systems having up to $100$ particles suggest that the effects we discuss should be observable in existing ultracold gas experiments. For example, the setup of~\cite{meinert_bloch_16} consists of a few (between one and three) impurities and about $60$ host atoms of cesium confined in a one-dimensional trap. The final state of the impurity accelerated by a constant external force has been measured. A modification of this setup in which the external force can be switched on and off would make the phenomena discussed in this Letter experimentally accessible.

\begin{acknowledgments}
We thank A.K.~Fedorov, V.~Kasper, A.~Maitra, Y.E.~Shchadilova, and A.~Sykes for careful reading of the manuscript. The work of OG, VC, and MBZ is part of the Delta ITP consortium, a program of the Netherlands Organisation for Scientific Research (NWO) that is funded by the Dutch Ministry of Education, Culture and Science (OCW). The work of OG was partially supported by Project 1/30-2015  ``Dynamics and topological structures in Bose-Einstein condensates of ultracold gases''  of the KNU  Branch Target Training at the NAS of Ukraine. EB acknowledges support by the grant 14-21-00158 from the Russian Science Foundation. OL acknowledges the support from the Russian Foundation for Basic Research under Grant No.\ 16-32-00669. The work of MBZ is supported by the grant ANR-16-CE91-0009-01.
\end{acknowledgments}

%


\end{document}



\title{Supplemental Material for ``The impact of the injection protocol on an impurity's stationary state''}

\author{Oleksandr Gamayun}
\affiliation{Instituut-Lorentz, Universiteit Leiden, P.O. Box 9506, 2300 RA Leiden, The Netherlands}
\affiliation{Bogolyubov Institute for Theoretical Physics, 14-b Metrolohichna str., Kyiv 03680, Ukraine}

\author{Oleg Lychkovskiy}
\affiliation{Skolkovo Institute of Science and Technology,
Skolkovo Innovation Center 3, Moscow  143026, Russia}
\affiliation{Steklov Mathematical Institute of Russian Academy of Sciences,
Gubkina str. 8, Moscow 119991, Russia}
\affiliation{Russian Quantum Center, Novaya St. 100A, Skolkovo, Moscow Region, 143025, Russia}

\author{Evgeni Burovski}
\affiliation{National Research University Higher School of Economics, 101000 Moscow, Russia}
\affiliation{Science Center in Chernogolovka,142432 Chernogolovka, Russia}

\author{Matthew Malcomson}
\affiliation{Physics Department, Lancaster University, Lancaster LA1 4YB, United Kingdom}

\author{Vadim V. Cheianov}
\affiliation{Instituut-Lorentz, Universiteit Leiden, P.O. Box 9506, 2300 RA Leiden, The Netherlands}

\author{Mikhail B. Zvonarev}
\affiliation{LPTMS, CNRS, Univ. Paris-Sud, Universit\'e Paris-Saclay, 91405 Orsay, France}

\date{\today}


\maketitle

\section{S1 Excitation spectrum through the Bethe Ansatz solution, and Fig.~2($\mathbf{a}$)}

In this section we review some properties of the excitation spectrum of the model, which are found by using the Bethe Ansatz solution. In particular, these properties help in understanding Fig.~2(a) of the Letter. We let the particle mass $m=1$ through the rest of the Supplemental Material in order to lighten notations.

We have
\begin{equation}
H|f_\gamma\rangle = E_{f_\gamma} |f_\gamma\rangle,
\end{equation}
where $H$ is the Hamiltonian~(1). Its eigenfunctions $|f_\gamma\rangle$, and the corresponding energies $E_{f_\gamma}$ are found by the Bethe Ansatz~\cite{mcguire_impurity_fermions_65}. The total momentum of the system is a good quantum number. For the gas with $N$ particles each $|f_\gamma\rangle$ can be uniquely identified by $N+1$ numbers $k_1,\ldots,k_{N+1}$ called quasi-momenta. These numbers are real for the repulsive impurity-gas interaction, $\gamma\ge 0$. The total momentum $Q$ and the energy $E_{f_\gamma}$ are obtained through the quasi-momenta as if the system consists of free particles:
\begin{equation}\label{QQ}
Q=\sum\limits_{j=1}^{N+1} k_j,
\end{equation} 
and
\begin{equation}
E_{f_\gamma} = \frac12 \sum\limits_{j=1}^{N+1} k^2_j.
\label{eq:energy}
\end{equation} 
That the model is interacting is encoded in the quantization conditions for the quasi-momenta. For the system on a ring of circumference $L$ and periodic boundary conditions, $k_j$ are given by a set of the non-linear equations (Bethe equations)
\begin{equation}\label{bethe1}
k_j = \frac{2\pi}{L}\left(n_j-\frac{\delta_j}{\pi}\right), \quad \delta_j =R\left(\Lambda-\frac{4\pi}{\gamma N}n_j\right), \quad j=1,\ldots,N+1.
\end{equation}
Here, $R(x)$ is a solution to the equation 
\begin{equation}\label{bethe2}
\cot R(x) = x + \frac{4R(x)}{\gamma N}, \quad \frac{\partial R(x)}{\partial x}<0, \quad R(-\infty)=\pi, \quad R(\infty)=0,
\end{equation}
$n_1,\ldots,n_{N+1}$ is a set of distinct integers, and the parameter $\Lambda$ is determined by Eq.~\eqref{QQ}. We have for Eq.~\eqref{bethe2}
\begin{equation}
R(x) = \frac{\pi}{2}-\arctan(x), \quad L\to\infty.
\label{approxdelta}
\end{equation}

The thermodynamic limit of the model~(1) leads to 
\begin{equation}\label{phase}
\frac{Q}{k_F}= \frac1{k_F} \frac{2\pi}{L}\sum\limits_{j=1}^{N+1} n_j -1+
\frac{1}{\alpha\pi} \left[
\left(\Lambda+\alpha\right)\arctan\left(\Lambda+\alpha\right)
-\left(\Lambda-\alpha\right)\arctan\left(\Lambda-\alpha\right)+\frac{1}{2}\ln \frac{1+(\alpha-\Lambda)^2}{1+(\alpha+\Lambda)^2}\right].
\end{equation}
for Eq.~\eqref{QQ}, and to	
\begin{equation}\label{Energy}
E_{f_\gamma} = \frac{1}{2} \sum\limits_{j=1}^{N+1} \left(\frac{2\pi}{L}n_j\right)^2 + E(\Lambda)
\end{equation}
for Eq.~\eqref{eq:energy}, where
\begin{equation}
\frac{E(\Lambda)}{E_F} = \frac{1}{\pi\alpha} - \left(\alpha^2+1-\Lambda^2\right)\frac{\arctan\left(\alpha-\Lambda\right)+\arctan\left(\alpha+\Lambda\right)}{2\pi\alpha^2} + \frac{\Lambda	}{2\pi \alpha^2} \ln\frac{1+(\alpha-\Lambda)^2}{1+(\alpha+\Lambda)^2}.
\end{equation}
Recall that $\alpha=2\pi/\gamma$, the gas density $\rho=N/L$, and the Fermi momentum $k_F=\pi\rho$. The set of the quantum numbers ($N$ is odd hereafter)
\begin{equation}
n_j = -\frac{N+1}{2}+j, \quad j=1,\ldots,N+1
\end{equation}
delivers the minimum of $E_{f_\gamma}$ (denoted as $E_\mathrm{min}$) at a given total momentum $0\le Q\le 2k_F$. Equations~\eqref{phase} and~\eqref{Energy} connect $E_\mathrm{min}$ and $Q$ through the parameter $\Lambda$. Resolving them numerically, we obtain the plots shown in Fig.~2(a) of the Letter.

\section{S2 Details of the derivation of Eqs.~(6) and (20)}

In this section we detail the derivation of Eqs.~(6) and (20). For this we exploit results and notations from Ref.~\cite{gamayun_impurity_Green_FTG_16}.

We write the Hamiltonian (1) in the mobile impurity reference frame:
\begin{equation}\label{HH}
H_\mathcal{Q} \equiv \mathcal{Q} H \mathcal{Q}^{-1} =  \frac12\left(Q-\sum\limits_{j=1}^N P_j\right)^2  +  \frac12\sum\limits_{j=1}^N P_j^2 + \gamma\rho\sum\limits_{j=1}^N \delta(x_j), \quad \mathcal{Q}=e^{ix_\mathrm{imp}\sum_{j=1}^N P_j}.
\end{equation}
Let us consider the matrix element $ \langle f_\gamma|P_\mathrm{imp}|f_\gamma\rangle $, which enters Eqs.~(5) and~(19). In the mobile impurity reference frame it is given by the derivative of the Hamiltonian~\eqref{HH} with respect to $Q$:
\begin{equation}	
\langle f_\gamma |P_\mathrm{imp}| f_\gamma\rangle = \langle f_\gamma |\mathcal{Q}^{-1} (Q-\sum\limits_{j=1}^N P_j) \mathcal{Q}| f_\gamma \rangle  = \langle f_\gamma |\mathcal{Q}^{-1} \frac{\partial H_\mathcal{Q}}{\partial Q} \mathcal{Q}| f_\gamma \rangle.
\end{equation}
Using the Hellmann-Feynman theorem we arrive at 
\begin{equation}\label{PP}
\langle f_\gamma |P_\mathrm{imp}| f_\gamma \rangle = \frac{\partial E_{f_\gamma}(Q)}{\partial Q}.
\end{equation}
This explains how Eq.~(20) is obtained from Eq.~(19). The explicit expression for the right hand side of Eq.~\eqref{PP} follows from Eq.~\eqref{Energy} and does not depend on the quantum numbers $n_j$:
\begin{equation} \label{PP2}
\mathcal{P}(\Lambda) \equiv \langle f_\gamma |P_\mathrm{imp}| f_\gamma \rangle  =
 \frac{\Lambda }{\alpha }+\frac{1}{2\alpha}\frac{\ln\frac{1+(\alpha -\Lambda )^2}{1+(\alpha +\Lambda )^2}}{\arctan(\alpha -\Lambda) + \arctan(\alpha +\Lambda)}.
\end{equation}
Combining Eqs.~\eqref{PP2} and \eqref{phase} for the minimum of the excitation spectrum, we get $v_\infty^m$ for $v_0<v_F$ shown in Fig.~1(b) of the Letter.

The eigenstates $|f_\gamma\rangle$ over which the sum in Eq.~(5) is taken are parametrized by the quantum numbers $n_1,\ldots,n_{N+1}$ and by the parameter $\Lambda$. We therefore write
\begin{equation}\label{sum}
\sum_{f_\gamma} \langle f_\gamma|P_\mathrm{imp} |f_\gamma\rangle |\langle f_\gamma | \mathrm{FS},Q \rangle|^2 = \sum_\Lambda \mathcal{P}(\Lambda) \sum_{n_1,\ldots,n_{N+1}} |\langle f_\gamma | \mathrm{FS},Q \rangle|^2.
\end{equation}
We further transform Eq.~\eqref{sum} by using the form-factor summation technique from Ref.~\cite{gamayun_impurity_Green_FTG_16}. Equations~(4.19), (6.7), and (6.33) of Ref.~\cite{gamayun_impurity_Green_FTG_16}, modified to our case straightforwardly, bring the right hand side of Eq.~\eqref{sum} to the desired form, given by Eq.~(6) of the Letter.

\section{S3 Details of the asymptotic and numeric analysis of Eq.~(6)}

In this section we demonstrate how Eqs.~(13) and (16) follow from Eq.~(6). We also explain how to evaluate Eq.~(6) numerically in order to get Fig.~1(a).

We have for Eq.~(6):
\begin{equation}
\det(\hat I +\hat V) = \det(\hat I +\hat V - \hat W)=1, \quad \gamma\to 0 \quad \textrm{or} \quad v_0\to\infty.
\end{equation}
Indeed, the kernels $V(q,q^\prime)$ and $W(q,q^\prime)$ are proportional to $\gamma$ in the $\gamma\to0$ limit. In the case $v_0\to\infty$ the main contribution to the integral over $x$ come from the vicinity of the point $x=0$, and the kernels vanish at $x=0$. Therefore we get
\begin{equation}\label{asdf}
\frac{v_\infty^{i}}{v_F} = \int_{-\infty}^\infty \frac{d\Lambda}{\pi} \frac{\mathcal{P}(\Lambda)}{1+(\alpha v_0/v_F -\Lambda)^2}, \quad \gamma\to 0 \quad \textrm{or} \quad v_0\to\infty.
\end{equation}
We take the integral and arrive at
\begin{equation}\label{asdf2}
\frac{v_\infty^{i}}{v_F} = \mathcal{P}(\alpha v_0/v_F), \quad \gamma\to 0 \quad \textrm{or} \quad v_0\to\infty.
\end{equation}
In the case $\gamma\to0$ and an arbitrary $v_0$ Eq.~\eqref{asdf2} leads to Eq.~(13) of the Letter, and in the case of $v_0\to\infty$ and an arbitrary $\gamma$ it leads to Eq.~(16) of the Letter.

We now turn to describing how we evaluate Eq.~(6) numerically. For a fixed $\Lambda$ and $x$ a single Fredholm determinant can be evaluated by the method described in Ref.~\cite{bornemann_10}. Using the Gauss-Legendre quadrature rule for $50$ points on the interval $[-1,1]$ the determinants are evaluated to a decent precision for any real $\Lambda$ and $x\lesssim 100$. However, 
$F(\Lambda,x)$ decays slowly with increasing $x$, and the integration over $x$ and $\Lambda$ leads to a significant cumulative error. In order to circumvent this problem we transform the integrand using its analytic properties before doing the numerics.

\begin{figure*}[ht]
\includegraphics[width=0.5\linewidth, clip=true,trim= 0 0 0 0]{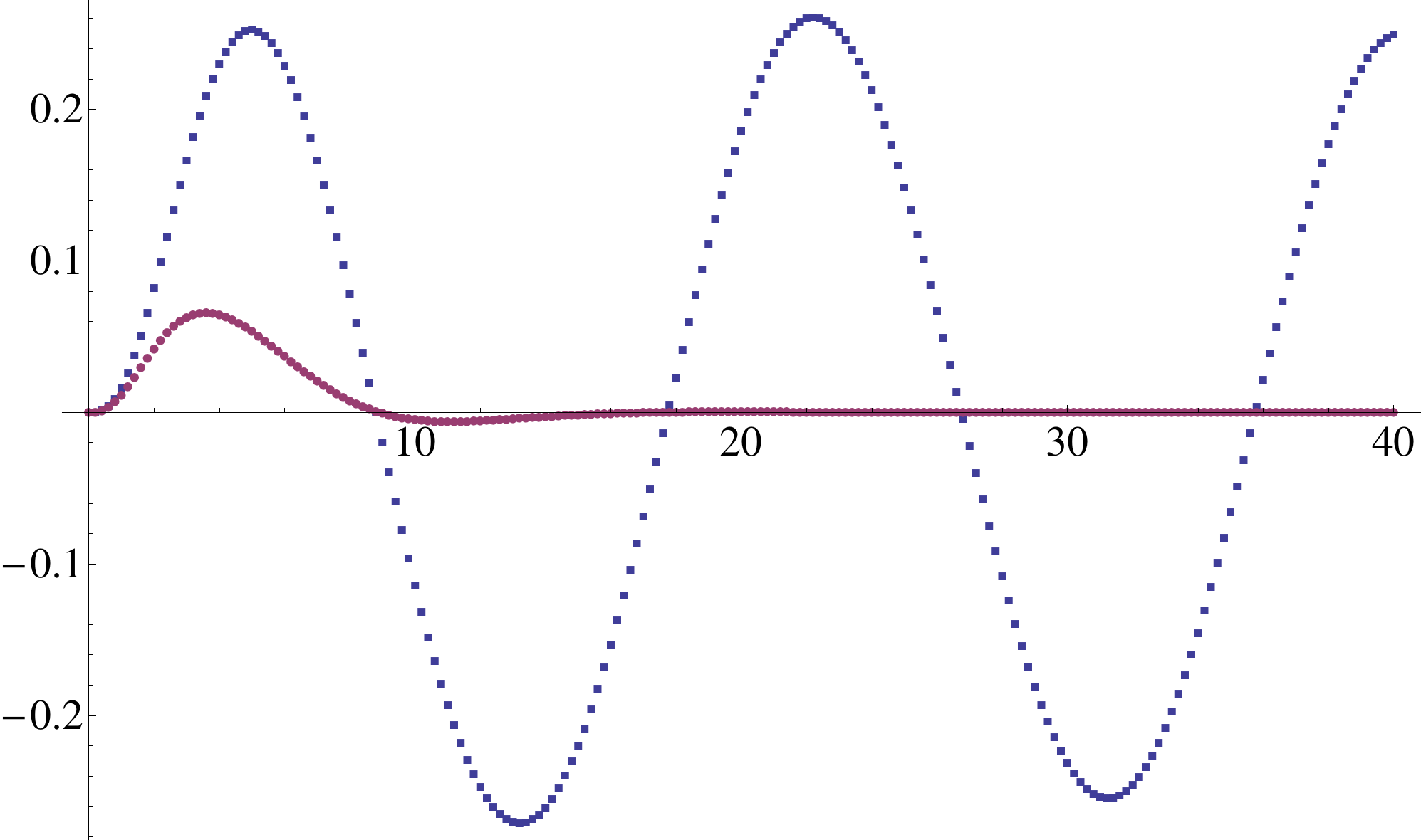}
\caption{\label{decay}
The blue dotted line shows $\tilde{F}(x,\alpha)$ as a function of $x$, and the purple line is for $\tilde{F}(x,\alpha+i)$. The decay rates for these two curves are remarkably different. The coupling strength $\gamma=3$.}
\end{figure*}

We present Eq.~(6) in the form 
\begin{equation} \label{v01}
v_\infty^i =v^{(0)} +v^{(1)},
\end{equation}
where $v^{(0)}/v_F $ is given by Eq. \eqref{asdf} and 
\begin{equation}\label{v1}
\frac{v^{(1)}}{v_F} = \frac{1}{ \pi i} \int\limits_{-\infty}^\infty d\Lambda\, \mathcal{P}(\Lambda) \int\limits_{0}^\infty dx \, \sin(xv_0/v_F) \tilde{F}(\Lambda,x),
\end{equation}
where
\begin{equation}\label{fL}
\tilde{F}(\Lambda,x) = [\det(\hat I+\hat{V})-1]h(x)+\det(\hat I+\hat{V}-\hat{W})-\det(\hat I+\hat{V}).
\end{equation}
We note that this function is analytic for the $\Lambda$ in the upper half-plane, and decays there as $1/\Lambda$. The 
function $\mathcal{P}(\Lambda)$ can be represented as
\begin{equation}
\mathcal{P}(\Lambda) =  \frac{ \Lambda }{\alpha}+\frac{1}{i \alpha }\frac{\ln\left(\frac{\Lambda-\alpha +i}{\Lambda +\alpha+i}\right)+\ln\left(\frac{\Lambda-\alpha -i}{\Lambda +\alpha-i}\right)}{\ln\left(\frac{\Lambda +\alpha+i}{\Lambda-\alpha +i}\right)-\ln\left(\frac{\alpha+\Lambda -i}{\Lambda-\alpha -i}\right)}.
\end{equation}
It has the branch-cut from $i-\alpha$ to $i+\alpha$. The jump across the branch-cut is 
\begin{equation}
\Delta_\mathcal{P}(y) \equiv  \mathcal{P}(y+i-i0) - \mathcal{P}(y+i+i0) = \frac{4\pi}{\alpha} \frac{\ln\left(\frac{y-\alpha+2i}{y+\alpha+2i}\right)}{\pi^2 +\left[\ln\left(\frac{y+\alpha+2i}{y-\alpha+2i}\right)+ \ln\left(\frac{\alpha-y}{\alpha+y}\right)\right]^2}, \quad y \in [-\alpha,\alpha].
\end{equation}
Therefore we can transform integral in Eq. \eqref{v1} as
\begin{equation}\label{vv1}
\frac{v^{(1)}}{v_F} = \frac{1}{ \pi i} \int\limits_{-\alpha}^\alpha dy\, \Delta_\mathcal{P}(y) \int\limits_{0}^\infty dx \, \sin(xv_0/v_F) \tilde{F}(y+i,x).
\end{equation}
The function $\tilde{F}(y+i,x)$ decays exponentially with increasing $x$, in the contrast to the function~\eqref{fL} for the real $\Lambda$. This is illustrated in Fig.~S\ref{decay}. Another advantage of Eq.~\eqref{vv1} is that the integral over $y$ is defined on the compact domain. Using Eq.~\eqref{vv1} we get the plots shown in Fig.~1(a) of the Letter.

\section{S4 The state $|\tilde f_\gamma\rangle$ for the adiabatic injection, and Eqs.~(19) and~(21)}

In this section we identify the state $|\tilde f_\gamma\rangle$ entering Eq.~(19), and prove Eq.~(21).

The first-quantized representation of the Fermi sea wave function is
\begin{equation}\label{initial}
|\mathrm{FS}\rangle = \frac{1}{\sqrt{N!L^N}}  \det_{1\le j,k \le N} (e^{i x_{j}q_{k}}), \quad q_j=  \frac{2\pi}{L}m_j, \quad m_j = -\frac{N-1}{2}+j-1, \quad j= 1,\dots N.
\end{equation}
It is an eigenfunction of the Hamiltonian \eqref{HH} at $\gamma=0$:
\begin{equation}\label{EEE}
H_\mathcal{Q}|\mathrm{FS}\rangle = \left( \frac{Q^2}{2} + \sum \limits_{j=1}^{N} \frac{q_j^2}{2} \right) |\mathrm{FS}\rangle, \quad \gamma=0.
\end{equation}
The eigenfunctions of the Hamiltonian~\eqref{HH} at arbitrary $\gamma$ read in the coordinate representation as the determinant of the $(N+1) \times (N+1)$ matrix
\begin{equation}\label{wavefunction}
|f_\gamma\rangle=
\frac{Y_{f_\gamma}}{\sqrt{N! L^N}} \begin{vmatrix}
e^{ik_1 x_1} &\dots & e^{ik_{N+1} x_1}\\
\vdots & \ddots & \vdots \\
e^{ik_1 x_N} &\dots & e^{ik_{N+1} x_N}\\
\nu(k_1) & \dots  & \nu(k_{N+1})\\
\end{vmatrix}, \quad \nu(k) = \frac{1}{\frac{2L}{\gamma N} k-\Lambda-i},
\end{equation}
where the factor $Y_{f_\gamma}$ ensures the normalization condition $|\langle f_\gamma| f_\gamma\rangle|=1$.

Let us examine Eqs.~\eqref{bethe1} and \eqref{bethe2} in the $\gamma\to 0$ limit. Representing $\delta_j$ from Eq.~\eqref{bethe1} as 
\begin{equation}
\delta_j = R\left[\frac{2L}{\gamma N}\left(k_* - \frac{2\pi}{L}n_j\right)\right], \quad k_*= \Lambda\frac{\gamma N}{2L}
\end{equation}
and using the asymptotic expansion for Eq.~\eqref{bethe2}
\begin{equation}
R(x) =\left\{ \begin{array}{lcl}
\pi & & x\le -\dfrac{4\pi}{\gamma N}\\
-\dfrac{\gamma N}{4}x & & -\dfrac{4\pi}{\gamma N}<x<0\\
0& & x>0
\end{array} \right. , \quad \gamma \to 0
\end{equation}
we get
\begin{equation}
\delta_j =\left\{ \begin{array}{lcl}
\pi & & k_*\le \dfrac{2\pi}L (n_j-1)\\
-\dfrac{L}{2} \left(k_* -\dfrac{2\pi}L n_j\right) & & \dfrac{2\pi}L (n_j-1) < k_* < \dfrac{2\pi}L n_j\\
0& & k_* \ge \dfrac{2\pi}L n_j
\end{array} \right. , \quad \gamma \to 0.
\end{equation}

Let us consider $0\le Q \le k_F$ (recall that we consider $Q\ge 0$ in order to simplify the notations). Since $Q$ is quantized we write
\begin{equation}
Q = \frac{2\pi}L \left( -\frac{N+1}{2}+M \right), \quad M=\frac{N+1}2,\ldots,N.
\end{equation}
We take the solution to the Bethe equations corresponding to the quantum numbers distributed as follows:
\begin{equation}
n_j = - \frac{N+1}2 + j,  \quad j=1,\ldots,N+1.
\end{equation}
We find
\begin{equation}
\delta_j =\left\{ \begin{array}{lcl}
\pi & & j=M+1,\ldots, N+1 \\
0& & j = 1,\ldots,M
\end{array} \right. , \quad \gamma \to 0,
\label{delta0}
\end{equation}
and
\begin{equation}
k_* =Q, \quad \{k_1,k_2,\dots,k_{N+1}\} = \{q_1,q_2,\dots,q_{M-1}, Q,Q,q_{M+1},\dots, q_N\} , \quad \gamma \to 0. \label{ktogammaslow}
\end{equation}
Note that $Q=q_M$. This way, the quasi-momenta satisfy Eq.~\eqref{QQ}, and the energy \eqref{eq:energy} has the value determined by Eq.~\eqref{EEE}. However, the $M$th and $M+1$th columns of the matrix in Eq.~\eqref{wavefunction} coincide in the $\gamma\to 0$ limit, and the analysis of the determinant requires solving Eqs.~\eqref{bethe1} and \eqref{bethe2} to the next to the leading order in $\gamma$.

It follows from Eqs.~\eqref{bethe1} and \eqref{bethe2} that
\begin{equation}
\cot\delta_j = \frac{2L}{\gamma N} (k_*-k_j), \quad j=1,\ldots,N+1. \label{bethe3}
\end{equation}
We seek the small $\gamma$ solution to Eqs.~\eqref{bethe3} assuming that
\begin{equation} \label{asympt}
\frac{k_*-k_M}{\gamma N} \to \infty, \quad \frac{k_*-k_{M+1}}{\gamma N} \to -\infty, \quad \textrm{as} \quad \gamma\to0.
\end{equation}
We write for the first subleading corrections in Eq.~\eqref{delta0}
\begin{equation}\label{deltacorrsmall}
\delta_j =\left\{ \begin{array}{lcl}
\dfrac{\gamma N}{2L(k_*-k_j)} & & j=1,\ldots, M \\
\pi+\dfrac{\gamma N}{2L(k_*-k_j)} & & j = M+1,\ldots,N+1
\end{array} \right. , \quad \gamma \to 0
\end{equation}
and in Eq.~\eqref{QQ}
\begin{equation} \label{sumas}
\frac1{k_*-k_M}+\frac1{k_*-k_{M+1}} + \sum_{\substack{j=1\\ j\ne M}}^N \frac1{k_*-q_j} =0, \quad \gamma\to 0.
\end{equation}
The first and the second terms in Eq.~\eqref{sumas} diverge in the $\gamma\to 0$ limit, and their sum is regular. The asymptotic expansion for the quasi-momenta satisfying Eqs.~\eqref{asympt}--\eqref{sumas} reads
\begin{equation}
k_M = q_M - \frac{\sqrt{\gamma N}}L , \quad
k_{M+1} = q_{M} + \frac{\sqrt{\gamma N}}L, \quad k_* = q_M + \frac{2\pi}{L}\epsilon, \quad\gamma\to0.
\end{equation}
Here, $\epsilon$ decays as a first power of $\gamma$ or faster, in the $\gamma\to0$ limit. Therefore, the function $\nu(k_j)$ defined in Eq.~\eqref{wavefunction} behaves as follows:
\begin{equation}
\nu(k_M) = -\nu_Q, \quad \nu(k_{M+1}) = \nu_Q, \quad \nu_Q = \frac{\sqrt{\gamma N}}2, \quad \gamma\to0,
\end{equation}
and decays as a first power of $\gamma$ or faster for $j\ne M$ and $M+1$. We thus get for the wave function~\eqref{wavefunction}
\begin{equation}
|f_\gamma\rangle =
\frac{Y_{f_\gamma}}{\sqrt{N!L^N}} \begin{vmatrix}
e^{iq_1 x_1} &\dots &e^{iq_Mx_1} &e^{iq_Mx_1} &\dots &e^{iq_{N} x_1}\\
\vdots & \ddots &\vdots &\vdots &\ddots &\vdots \\
e^{iq_1 x_N} &\dots  &e^{iq_Mx_N}& e^{iq_Mx_N} & \dots& e^{iq_{N} x_N}\\
0 & \dots  & -\nu_Q & \nu_Q &\dots &0\\
\end{vmatrix}, \quad \gamma\to0.
\end{equation}
The determinant can be taken explicitly, and, using Eq.~\eqref{initial}, we arrive at
\begin{equation}
|f_\gamma\rangle = (-1)^{M+1}2\nu_Q Y_{f_\gamma} |\mathrm{FS}\rangle, \quad \gamma\to0. \label{fgammalim}
\end{equation}
The normalisation factor $Y_{f_\gamma}$ is given explicitly by Eq.~(5.22) in Ref.~\cite{gamayun_impurity_Green_FTG_16}:
\begin{equation}\label{norm}
|Y_{f_\gamma}|^{-2} = \prod_{j=1}^{N+1} \left| 1+ \frac4{\gamma N} \nu_-(k_j)\nu_+(k_j)\right| \left| \sum_{j=1}^{N+1} \frac{\nu_-(k_j)\nu_+(k_j)}{1+ \frac4{\gamma N} \nu_-(k_j)\nu_+(k_j)} \right|, \quad \nu_{\pm}(k) = \frac{1}{\frac{2L}{\gamma N} k-\Lambda \pm i}.
\end{equation}
Therefore
\begin{equation}
|Y_{f_\gamma}|^{-2} = (2\nu_Q)^2, \quad \gamma\to 0.
\end{equation}
Substituting this formula into Eq.~\eqref{fgammalim} we complete the proof of Eq.~(18) of the Letter for $Q$ smaller than $k_F$. Note that this proof is valid for an arbitrary $N$.

Let us now turn to the case $Q>k_F$. The arguments presented earlier in the section are modified as follows. We write
\begin{equation}
Q= \frac{2\pi}L \left(-\frac{N+1}2 +M\right), \quad M\ge N+1 \label{Qlarge}
\end{equation}
and take the solution to the Bethe equations corresponding the following distribution of the quantum numbers:
\begin{equation} \label{njlarge}
n_j = -\frac{N+1}2+j, \quad j=1,\ldots,N, \quad \textrm{and} \quad n_{N+1}= -\frac{N+1}2+M+1.
\end{equation}
 We have 
\begin{equation}
\delta_j =\left\{ \begin{array}{lcl}
\pi & & j= N+1 \\
0& & j = 1,\ldots,N
\end{array} \right. , \quad \gamma \to 0,
\label{delta0fast}
\end{equation}
and
\begin{equation}\label{ktogammafast}
\{k_1,k_2,\dots,k_{N+1}\} = \{q_1,q_2,\dots, q_N,Q\} , \quad \gamma \to 0 
\end{equation}
instead of Eqs.~\eqref{delta0} and~\eqref{ktogammaslow}, respectively. Equation~\eqref{deltacorrsmall} changes accordingly, and in place of Eq.~\eqref{sumas} we have
\begin{equation} \label{sumasfast}
\frac1{k_*-k_{N+1}} + \sum_{j=1}^N \frac1{k_*-q_j} =0, \quad \gamma\to 0.
\end{equation}
Note that $k_*\ne Q$ at $\gamma=0$ for finite $N$, and it tends to $Q$ in the $N\to\infty$ limit. The sum over $j$ in Eq.~\eqref{sumasfast} diverges linearly with $N$:
\begin{equation}
\frac1{k_*-k_{N+1}} =-\frac{N}{2k_F} \ln\frac{Q+k_F}{Q-k_F}, \quad N\to\infty \quad \text{after} \quad \gamma\to 0.
\end{equation}
The function $\nu(k_j)$ behaves as follows
\begin{equation} \label{nun+1}
\nu(k_{N+1}) = \frac{\gamma N}{4\pi} \ln\frac{Q+k_F}{Q-k_F}, \quad N\to\infty \quad \text{after} \quad \gamma\to 0,
\end{equation}
and
\begin{equation}\label{nuj}
\nu(k_j) = \frac{\gamma}{2\pi}\frac{k_F}{q_j-Q}, \quad j=1,\ldots,N, \quad N\to\infty \quad \text{after} \quad \gamma\to 0.
\end{equation}
We see that the scaling with $\gamma$ is the same for any $j=1,\ldots,N+1$. The function $|f_\gamma\rangle$ defined by Eq.~\eqref{wavefunction} converge to the function~\eqref{initial} only if the $N\to\infty$ limit is taken after the $\gamma\to 0$ limit. Indeed, $\nu(k_j)$ grows linearly with $N$ at $j=N+1$, and stays constant for all other values of $j$. We thus write for the wave function~\eqref{wavefunction}
\begin{equation}
|f_\gamma\rangle =\frac{Y_{f_\gamma}}{\sqrt{N!L^N}} \begin{vmatrix}
e^{iq_1 x_1} &\dots & e^{iq_N x_1}& e^{iQ x_1}\\
\vdots & \ddots & \vdots &\vdots \\
e^{iq_1 x_N} &\dots &  e^{iq_N x_N} & e^{iQ x_N}\\
0 & \dots  &0 & \nu(k_{N+1})\\
\end{vmatrix} = Y_{f_\gamma}\nu(k_{N+1})|\mathrm{FS}\rangle, \quad N\to\infty \quad \text{after} \quad \gamma\to 0.
\end{equation}
Substituting Eqs.~\eqref{nun+1} and \eqref{nuj} into Eq.~\eqref{norm} we find that
\begin{equation}
|Y_{f_\gamma}\nu(k_{N+1})|=1, \quad N\to\infty \quad \text{after} \quad \gamma\to 0.
\end{equation}
This completes the proof of Eq.~(18) of the Letter for $Q$ larger than $k_F$.

Substituting Eqs.~\eqref{Qlarge} and~\eqref{njlarge} into Eq.~\eqref{phase} we find that $\Lambda=\infty$ in the thermodynamic limit. Taking the $\Lambda\to\infty$ limit in Eq.~\eqref{PP2} we obtain Eq.~(21) of the Letter.

\section{S5 Effective mass $m_*$, and Fig.~2($\mathrm{b}$)}

In this section we explain how to calculate the effective mass $m_*$ of the polaron, and how to obtain the plots in Fig.~2(b).

The effective mass $m_*$ is determined by expanding the minimal energy of the excitations, $E_{\mathrm{min}}$, at $Q=0$:
\begin{equation}
E_{\mathrm{min}}(Q) - E_{\mathrm{min}}(0) = \frac{Q^2}{2m_*}, \quad Q\to 0.
\end{equation}
The analytic formula follows from Eqs.~\eqref{phase} and~\eqref{Energy} readily
\begin{equation}\label{eq:m*}
m_* = \frac{2}{\pi} \frac{\left( \arctan \alpha \right)^2} {\arctan \alpha - \alpha \left(1+ \alpha^2 \right)^{-1} }.
\end{equation}
The effective mass increases with increasing impurity-gas repulsion. For $\gamma \ll 1$ it deviates from the gas particle mass only slightly,
\begin{equation}\label{SmallM}
m_* = 1+\frac{\gamma^2}{\pi^4} + \cdots,\quad \gamma\to 0,
\end{equation}
while for $\gamma\gg 1$ it grows linearly with $\gamma,$ 
\begin{equation}\label{LargeM}
m_* = \frac{3\gamma}{2\pi^2}+ \cdots, \qquad \gamma\to\infty.
\end{equation}

To get the plots shown in Fig.~2(b) of the Letter we expand Eq.~(20) at $Q=0$:
\begin{equation}
\left.\frac{v_\infty^m}{v_0}\right|_{v_0=0} = \frac{1}{m_*}. \label{eq:vaQ0}
\end{equation}
We then expand Eq.~(6) at $Q=0$:
\begin{equation}
\left.\frac{v_\infty^i}{v_0}\right|_{v_0=0} = -i \int_{-\infty}^\infty \frac{d\Lambda}{\pi} \mathcal{P}(\Lambda) \int_0^\infty dx\, xF(\Lambda,x). \label{eq:viQ0}
\end{equation}
Therefore
\begin{equation}
\left. \frac{v_\infty^i}{v_\infty^m}\right|_{v_0=0} = m_* \left.\frac{v_\infty^i}{v_0}\right|_{v_0=0}. \label{eq:vratio}
\end{equation}
Numerical evaluation of Eqs.~\eqref{eq:viQ0} and~\eqref{eq:vratio} leads to the dotted and solid curves in Fig.~2(b) of the Letter, respectively.

The dashed horizontal line in Fig.~2(b) of the Letter corresponds to the $\gamma\to\infty$ limit of Eq.~\eqref{eq:vratio}. Let us use the representation~\eqref{v01}. The function $v^{(0)}$ decays faster than $1/\gamma$ in the $\gamma\to\infty$ limit, and the branch-cut in the uppper half-plane in $\mathcal{P}(\Lambda)$ transforms into a pole at $\Lambda=i$: 
\begin{equation}
\mathcal{P}(\Lambda) = \frac{2\alpha\Lambda}{3(1+\Lambda^2)}, \quad \gamma\to\infty.
\end{equation} 
We find from Eq.~(12)
\begin{equation}
e(q) = - \frac{e^{iqx}}{2i}, \quad h = \delta(x), \quad \textrm{for} \quad \Lambda=i \quad \textrm{and} \quad \gamma\to\infty.
\end{equation} 
This way we get
\begin{equation}\label{vvvv1}
\frac{v_\infty^i}{v_F} = \frac{2\alpha}{3}\int_0^\infty dx\, \sin(v_0x)[\det(\hat I +\hat V_\infty - \hat W_\infty) - \det(\hat I +\hat V_\infty)], \quad \gamma\to\infty,
\end{equation}
where
\begin{equation}
V_\infty (q,q^\prime) = -\frac1\pi \frac{\sin[\frac{x}2(q-q^\prime)]}{q-q^\prime}, \quad
W_\infty (q,q^\prime) = \frac1{4\pi} e^{ixq/2}e^{ixq^\prime/2}.
\end{equation}
Combining Eqs.~\eqref{LargeM}, \eqref{eq:vratio} and~\eqref{vvvv1} we arrive at
\begin{equation}
\left.\frac{v_\infty^i}{v_\infty^m}\right|_{v_0=0} = \frac{2}{\pi} \int_0^\infty dx\, x[\det(\hat I+\hat V_\infty - \hat W_\infty) - \det(\hat I+\hat V_\infty)], \qquad \gamma\to \infty.
\end{equation}
Numerical evaluation of the Fredholm determinants gives
\begin{equation}
\int_0^\infty dx\, x[\det(\hat I + \hat V_\infty - \hat W_\infty)- \det(\hat I + \hat V_\infty)] = 0.5726\ldots.
\end{equation}
Therefore
\begin{equation}
\left.\frac{v_\infty^i}{v_\infty^m}\right|_{v_0=0} = 0.3645\ldots, \qquad \gamma\to\infty.
\end{equation}

Note that to get the small $\gamma$ expansion of Eq.~\eqref{eq:viQ0} is a surprisingly difficult task. We were unable to get the first subleading term so far. We, however, have results from Ref.~\cite{gamayun_kinetic_impurity_TG_14}, obtained by a perturbation theory. Letting $\eta=1$ in Eq.~(40) therein, and expanding the resulting expression at $p_0=0$ we get
\begin{equation}
\left.\frac{v_\infty^i}{v_0}\right|_{v_0=0} = 1 - \frac{2\gamma^2}{\pi^4}, \quad \gamma\to 0. 
\end{equation}
Therefore
\begin{equation}
\left.\frac{v_\infty^i}{v_\infty^m}\right|_{v_0=0} = 1 - \frac{\gamma^2}{\pi^4}, \quad \gamma\to 0. 
\label{eq:viag0}
\end{equation}
